\documentclass[prl,aps,showpacs,twocolumn,eqsecnum,a4paper,ltxgrid]{revtex4}

\usepackage{graphicx}
\usepackage{mhchem}
\usepackage[sort&compress]{natbib}
\usepackage{amssymb}
\usepackage{booktabs}
\usepackage[utf8]{inputenc}
\usepackage[english]{babel}
\usepackage{amsfonts}
\usepackage{amsmath}
\usepackage{array}
\usepackage{multirow}
\usepackage{color}
\usepackage{subdepth}
\usepackage{babel}
\usepackage{todonotes}
\usepackage{chngcntr}
\usepackage{epsfig}

\makeatletter

\newcommand{\VB}{V_{\mathrm{B}}}
\newcommand{\VS}{V_{\mathrm{S}}}
\newcommand{\VSt}{\tilde{V}_{\mathrm{S}}}
\newcommand{\VSB}{V_{\mathrm{S-B}}}
\newcommand{\Vm}{V_{\mathrm{mean}}}
\newcommand{\Eq}[1]{Eq.~(\ref{#1})}

\newcommand{\Fig}[1]{Fig.~\ref{#1}}
\newcommand{\CL}{CL}

\newcommand{\diff}{\mathrm{d}}
\newcommand{\A}{\mathcal{A}}
\newcommand{\B}{\mathcal{B}}
\newcommand{\Ad}{\A_\mathrm{D}}
\newcommand{\Alp}{\A_\mathrm{LP}}
\newcommand{\Anlp}{\A_\mathrm{NLP}}
\@ifundefined{textcolor}{}
{%
 \definecolor{BLACK}{gray}{0}
 \definecolor{WHITE}{gray}{1}
 \definecolor{RED}{rgb}{1,0,0}
 \definecolor{GREEN}{rgb}{0,1,0}
 \definecolor{BLUE}{rgb}{0,0,1}
 \definecolor{CYAN}{cmyk}{1,0,0,0}
 \definecolor{MAGENTA}{cmyk}{0,1,0,0}
 \definecolor{YELLOW}{cmyk}{0,0,1,0}
}

\usepackage{blindtext}
\makeatother

\counterwithout{equation}{section}

\begin{document}

\title{On the Applicability of the Caldeira-Leggett Model to
Condensed Phase Vibrational Spectroscopy}

\author{Sergei D. Ivanov}
\author{Fabian Gottwald}
\email{fabian.gottwald@uni-rostock.de}
\author{Oliver K\"uhn}
\address{Institute of Physics, University of Rostock, Universit\"atsplatz 3, 18055 Rostock, Germany}

\date{\today}

\begin{abstract}
Formulating a rigorous system-bath partitioning approach remains an open issue.
In this context the famous Caldeira-Leggett  model that enables quantum and classical treatment of Brownian motion on equal footing
has enjoyed popularity.
Although this model is by any means a useful theoretical tool, its 
ability to describe anharmonic dynamics of real systems is often taken for granted.
In this Letter we show that the mapping between a molecular system under study and the model cannot be established
in a self-consistent way, unless the system part of the potential is taken effectively harmonic.
 Mathematically, this implies that the mapping is not invertible.
This `invertibility problem' is not dependent on the peculiarities of particular molecular systems under study
and is rooted in the anharmonicity of the system part of the potential.

\end{abstract}


\maketitle
\paragraph{Introduction and Theory.}
Model systems play an important role for our understanding of complex many-body dynamics. 
Reducing the description to a few parameters can not only ease the interpretation,
but enable the identification of key properties~\cite{may11}.
In condensed phase dynamics the spin-boson~\cite{leggett87_1} and
Caldeira-Leggett (CL)~\cite{Caldeira-PRL-1981,Caldeira-AP-1983}  models
have been the conceptual backbone of countless studies~\cite{grabert88_115}.
The latter has been extended to describe linear and non-linear spectroscopy within the second-order cumulant approximation,
termed multi-mode Brownian oscillator model in this context~\cite{Mukamel1995,Tanimura-PRE-1993}. 
It has become a popular tool for assigning spectroscopic signals in the last two decades.

The \CL\ model comprises an arbitrary system (coordinates $x$, potential $\VS(x)$), 
which is bi-linearly coupled to a bath of harmonic oscillators
(coordinates $Q_i$, potential $\VB(\{Q_i\})$) via a system-bath coupling potential $\VSB(x,\{Q_i\})$~\cite{Caldeira-PRL-1981}.
Later Caldeira and Leggett extended the model to an arbitrary function of system coordinates in the coupling and motivated the linearity of the coupling on the bath side~\cite{Caldeira-AP-1983}. Here,  we limit ourselves to the bi-linear version for the reasons that will become apparent later.
Restricting ourselves to a one-dimensional case yields~\cite{grabert88_115}
\begin{equation}
\label{eq:CL Hamiltonian}
\VB(\{Q_i\})+\VSB(x,\{Q_i\})\equiv \sum_{i}\frac{1}{2}\omega_{i}^{2}Q^2_{i}-\sum_{i}g_{i} Q_i x
\end{equation}
with the bath frequencies $\omega_{i}$, unit bath masses, and the coupling strengths $g_{i}$.
Often the square on the right hand side of \Eq{eq:CL Hamiltonian} is completed
causing thereby a harmonic correction to the system potential $\VS(x)$: $\VSt(x)\equiv \VS(x) - \sum_i g_i^2/(2 \omega_i^2) x^2$.
%
We stress, that omitting this harmonic correction is not self-consistent, see Supplement.
The \CL\ model imposes no restrictions on the form of the system potential, apart from the presence
of the aforementioned trivial correction.

The advantage of this model is the possibility to
derive from it a simple reduced equation of motion termed generalized Langevin equation (GLE), that reads
\begin{equation}
\label{eq:GLE}
\dot{p}(t) = -\frac{\partial \VSt(x)}{\partial x}-\intop_{0}^{t}\diff \tau\,\xi(t-\tau)p(\tau) + R(t)
\enspace ,
\end{equation}
where the bath is reduced to non-Markovian dissipation and fluctuations
represented by the memory kernel $\xi(t)$ and the stochastic force $R(t)$, respectively.
The former is connected via a cosine Fourier transform to the spectral density
\begin{equation}
 \label{eq:Spectral Density}
J(\omega)=\sum_{i}\frac{g_{i}^{2}}{\omega_{i}^{2}}\delta(\omega-\omega_{i})
\end{equation}
and the latter is a Gaussian process with zero-mean.
Importantly, the derivation can be performed both in classical~\cite{Mukamel1995,ZwanzigBook2001} and 
quantum~\cite{Ford-JSP-1987,Ford-PRA-1988} domains.
For the present purpose we would limit ourselves to the classical GLE, whose derivation from the \CL\ model requires no
further approximations.
In this case, the fluctuating force is connected to the memory kernel by means of the so-called fluctuation-dissipation theorem (FDT)
\begin{equation}
\label{eq:FDT}
\left\langle R(0)R(t)\right\rangle =mkT\xi(t)
\enspace ,
\end{equation}
which establishes the canonical ensemble with the temperature $T$.
We note that according to Ford \textit{et al.}~\cite{Ford-JSP-1987}: ``\ldots \textit{if} there is a universal description,
then it must be of the form we have obtained [that is \Eq{eq:GLE}]''.

The tremendous simplicity of the \CL\ model is, thus, that it reduces the system-bath interaction, $\VSB(x,\{Q_i\})$ to 
a single spectral density.
However, a real system Hamiltonian might very well differ from the \CL\ one.
In practical applications one, therefore, tries to find a mapping between the two in order to make full use of the model's simplicity.
Importantly, if such a mapping can be established, then the full quantum-mechanical treatment of the bath can be performed analytically
via the Feynman-Vernon influence functional~\cite{Feynman-AP-1963,Caldeira-PhysicaA-1983} without any further approximations.
Moreover, numerically exact hierarchy type equation of motion approaches
are essentially based on the \CL\ model~\cite{Dijkstra-PRL-2010,Suess:2014gz}.
Further, it is ideally suited for numerical methods that solve the Schr\"odinger equation in many dimensions~\cite{giese06_211}.
The \CL\ model can also be taken as a starting point for quantum-classical hybrid simulations, that is by treating only the usually
low-dimensional system part quantum-mechanically and the bath (semi)classically~\cite{egorov99_5238}.
Finally, the machinery for a purely classical treatment by means of a GLE is provided by the method of colored noise thermostats~\cite{Ceriotti2009a,Ceriotti2010}.
It is thus the \CL\ model that establishes a unified framework
for a quantum-classical comparison of dynamical properties in condensed phase systems provided that the real system can be mapped onto this model.
We note in passing that various models based on a coupling that is non-linear in system coordinates exist,
such as, e.g., the so-called square-linear \CL\ model~\cite{kuhn03:2155,Tanimura2009}.
They, however, correspond to different GLE types, which do not lead  to any known mapping, see Supplement, and are, thus, not considered here.
This puts forward the central question of this Letter, i.e.\ whether such a mapping onto the bi-linear \CL\ model exists for an arbitrary system.
Surprisingly, the amount of information concerning the applicability of this tremendously popular model is scarce.
Makri argued for the applicability if the system-bath coupling can be treated in the linear response regime~\cite{makri99:2823}.
Goodyear  and Stratt derived the model in the short time limit introducing so-called instantaneous normal modes~\cite{Goodyear1996}.
However, the bath statistics turned out to deviate from the predictions of the \CL\ model at longer times~\cite{Goodyear1996,Tuckerman1993a}.

In order to approach the central question, we discuss three possible ways for constructing a mapping onto the \CL\ model.
First, one might assume that the real bath can be either exactly or approximately described
by the \CL\ Hamiltonian in a certain coordinate system.
This case corresponds to a direct representation of the bath in a harmonic form, \Eq{eq:CL Hamiltonian}.
The system potential $\VS(x)$ would be then the actual system potential as given, for instance, by an explicit force field;
note the presence of the aforementioned harmonic correction in $\VSt(x)$.
%
Another possibility
is to rigorously derive the GLE from first principles for an arbitrary open system.
If the resulting equation coincided with that derived from the \CL\ model, it would then serve as an indirect proof of the model itself.
Here, Mori's projection operator techniques are commonly involved to derive a GLE-type equation from the
general Hamilton equations of motion, expressing noise and dissipation formally as  projected quantities~\cite{Mori1965,ZwanzigBook2001,Cubero-JCP-2005}.
The projection can be performed either in a linear or a non-linear way.
%
The second mapping is based on the linear projection, which
yields the GLE \Eq{eq:GLE} and the FDT \Eq{eq:FDT}
without any further approximations.
The system potential becomes effectively harmonic, i.e.\ $\VSt(x)=1/2m\tilde{\omega}^2 x^2$ at the price of projecting all the system anharmonicity into the bath.
The effective harmonic frequency follows as $\tilde{\omega}^2 = kT/\langle x^2 \rangle$ with $\langle \ldots \rangle$ denoting canonical averaging.
The third route employs a non-linear projection.
Here, the system potential is given by the potential of the mean force, $\Vm(x)$,
felt by the system at coordinate $x$, averaged over all bath coordinates
$Q$~\cite{Kawai2011,ZwanzigBook2001,Lange2006}.
In this formalism the memory kernel carries a functional dependence on the system's positions
and momenta and the fluctuating force has more complicated statistical
properties than given by the FDT in \Eq{eq:FDT}~\cite{Kawai2011}.
This functional dependence is hardly tractable
and an apparent simplification is to linearize
the kernel with respect to positions and momenta.
In this case the contribution proportional to $x$ vanishes and the resulting integral coincides with that in \Eq{eq:GLE}~\cite{Kawai2011}.

These three routes provide us with the three versions of the GLE that are parametrized from explicit MD simulations
upon inverting the equation
\begin{equation}
\label{eq:extraction}
\dot{C}_{pp}(t)=C_{pF}(t)-\intop_{0}^{t} \diff \tau \xi(t-\tau)C_{pp}(\tau)
\end{equation}
 to obtain the memory kernel $\xi(t)$, where $C_{pp}(t)$ and $C_{pF}(t)$
are the momentum--momentum and momentum--system force time-correlation functions, see Supplement.
Thus, the mapping between the explicit system under study and the \CL\ model can be formulated as the mapping
$\A: \{ C_{pp}(t);C_{pF}(t) \} \mapsto \xi(t)$.
The particular form of the system force, that is, the particular form of $\VSt(x)$, specifies the GLE employed and thus the 
mapping $\A$, referred to as $\Ad$, $\Alp$ and $\Anlp$ for the direct mapping,
mapping based on linear and on non-linear projection technique, respectively.
We note again that the latter mapping requires a linearization of the memory kernel.
In order to verify a mapping $\A$ and thus to deduce the applicability of the \CL\ model,
we compute the linear vibrational spectra from the GLEs and compare them against their 
explicit MD counterparts.
The match between the spectra describes thus the success of the system-bath model, see Fig.~S5 in the Supplement.

\paragraph{Models and Methods.}

We have employed three hydrogen-bonded systems: HOD in H$_2$O, OH in H$_2$O, and the ionic liquid
$[\mathrm{C}_{2}\mbox{mim}][\mbox{NTf}_{2}]$  at room temperature. These examples have been chosen since H-bonding is known to come along with a considerable anharmonicity in the potential energy surface (PES).  In particular,  HOD in H$_2$O or D$_2$O is among the most studied hydrogen-bonded systems~\cite{stenger01_027401} and the  comparison with the OH ion is used to
elucidate the importance of the intramolecular forces. The ionic liquid represents a system, where, in addition to moderate H-bonding,
there exists a strong Coulomb interaction between the ion pairs~\cite{roth12_105026}.

The HOD/OH in H$_2$O systems are comprised of 466 molecules in a periodic box with the length of $2.4\,$nm
interacting according to the force field taken from Ref.~\cite{Paesani-JCP-2010}.
In the latter case the HOD molecule is substituted by the diatomic OH fragment.
The ionic liquid  simulations were carried out in a periodic box with the length of $4.5\,$nm containing 216 ionic pairs and the force field described in Ref.~\cite{Koeddermann2007}.
All explicit molecular dynamics (MD) simulations were performed with the GROMACS program package (Version~4.6.5)~\cite{GROMACS}.
The spectra were computed for the OH stretch in the HOD/OH molecules and for  the C$(2)-$H stretch of the imidazolium ring~\cite{Koeddermann2007} in the ionic liquid, see Supplement for details.
%
In the latter case the potential for the C-H
stretching motion was re-parametrized to a Morse potential using DFT-B3LYP calculations~\cite{roth12_105026}.
Note that in the spirit of the system-bath treatment the O-H bond lengths were  taken as the  respective system coordinates, $x$, whereas all other degrees of freedom constituted the bath.
We employed the  ``standard protocol'' for calculating IR spectra, that is, a set of
NVE trajectories, each $6\,$ps long (time step $0.1\,$fs), was started from
uncorrelated initial conditions sampled from an NVT ensemble. 
The $C_{pp}$ functions for the system coordinate
were Fourier-transformed to yield the vibrational spectra~\cite{Ivanov-PCCP-2013};
note that these spectra are proportional to the absorption spectra in the  one-dimensional case.
%
In order to achieve convergence, 1000 trajectories for the considered stretching motion  were employed, see Supplement for further details.
As a numerical solver for the
GLE we adopted the method of colored noise thermostats~\cite{Ceriotti2009a,Ceriotti2010}.
The simulation setup for GLE simulations was the same as that for the explicit ones.

\begin{figure}[t]
\includegraphics[width=0.99\columnwidth]{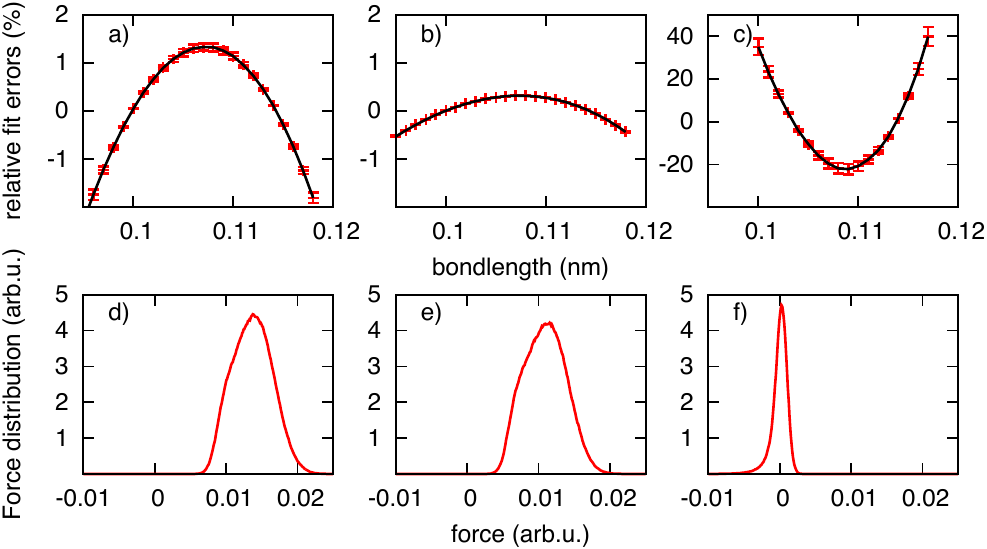}
\caption{\label{fig:MBO-check}
Panels a)-c): averaged $r(x)$,  see \Eq{eq:lin-fit-dev};
 d)-f): distribution functions of the forces corresponding to the noise term $R(t)$.
Left panels correspond to  HDO in bulk water; middle: to OH in bulk water; right: to the ionic liquid $[\mathrm{C}_{2}\mbox{mim}][\mbox{NTf}_{2}]$.
Note different scales on the $y$-axes.
}
\end{figure}

\paragraph{Results.}

Following the three routes described  above, we start with checking whether the direct mapping exists,
which implies that the actual Hamiltonian of the system studied has the form of \Eq{eq:CL Hamiltonian}.
The linearity of the coupling on the system side can be verified via explicit MD simulations in a straightforward manner.
Here, we sampled $5000$ uncorrelated configurations and varied the system coordinate, $x$,
 within the range accessible due to its thermal fluctuations
keeping all bath coordinates fixed, see Supplement.
The system-bath coupling $V_{\mathrm{S-B}}(x)$ probed this way was least-squares fitted to linear functions $f(x)=ax+b$.
In order to quantify the fit error with respect to a reasonable scale, we considered the relative deviation 
\begin{equation}
\label{eq:lin-fit-dev}
r(x)=\frac{f(x)-V_{\mathrm{S-B}}(x)}{ |V_{\mathrm{S-B}}(x_i)-V_{\mathrm{S-B}}(x_f)|}
\enspace ,
\end{equation}
with $x_i^{}/x_f^{}$ being the initial/final values of the probed range, respectively.
The results averaged over the chosen dataset are depicted in the upper panels of \Fig{fig:MBO-check} with the error bars given in red.
It becomes obvious that excellent fit results can be obtained for the two aqueous systems (panels a and b)
 where the relative deviations are below $2\,$\%.
In contrast, the averaged fit errors for the ionic liquid (panel c) are in the range of $50\mathrm{\%}$ on the scale of the overall change of the potential on
the accessible interval.
Hence, we can conclude that for aqueous systems the coupling $V_{\mathrm{S-B}}$ is linearly dependent on the bond length, $x$, as it is assumed in the \CL\ model, whereas it is by no means true for the ionic liquid considered.

\begin{figure}[tb]
\begin{centering}
\includegraphics[width=0.99\columnwidth]{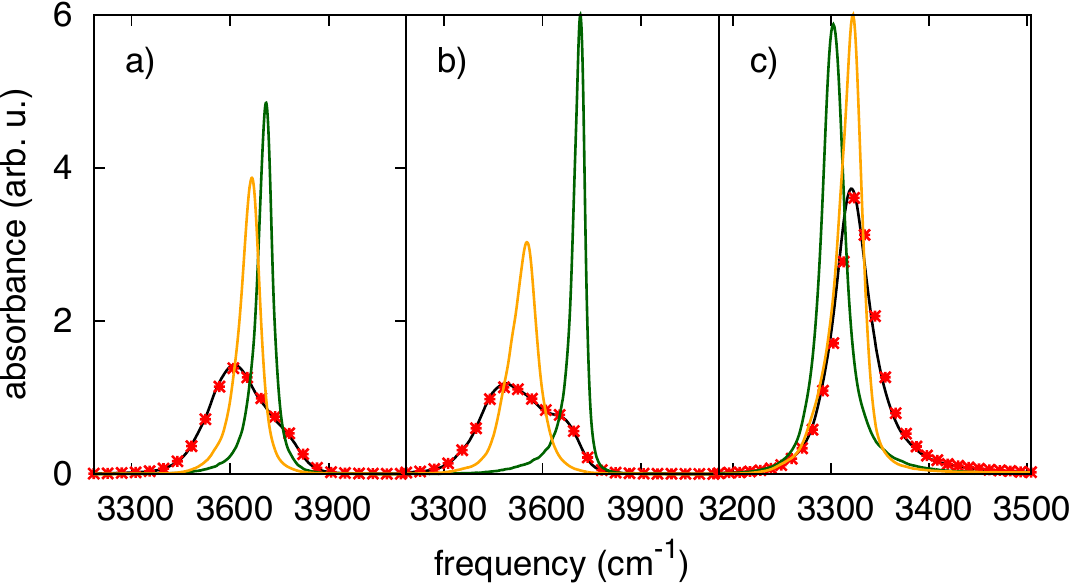}
\caption{\label{fig:spectra}
Explicit MD and GLE spectra for
a) HDO in bulk water,
b) OH in bulk water
 and 
c) the ionic liquid 
(black: explicit MD result;
red stars: $\Alp$;
green: direct mapping;
range: linearized $\Anlp$. See text for definitions.
}
\end{centering}
\end{figure}

Given the fact that the CL model is strictly equivalent to the GLE, the assumptions concerning the nonlinearity of the coupling on the bath side and the harmonicity of the bath, 
being hardly separable, together imply Gaussian statistics of the noise.
The Gaussian statistics can be easily probed via explicit MD simulations by calculating the distribution of the 
environmental forces exerted on the system coordinate, which is kept fixed to remove the dissipative contributions,
see Supplement.
The resulting distributions shown in the lower panels of \Fig{fig:MBO-check} are clearly asymmetric and centered at positive values 
for the two aqueous systems (panels d and e).
The distribution for the ionic liquid (panel f) is centered almost at zero and is only slightly skewed, thus looking
much more similar to a Gaussian.
To summarize, checking the linearity on the system side suggests that the model should perform well for aqueous systems 
and badly for the ionic liquid, whereas the corresponding results for the linearity on the bath side imply the opposite.
Another test for the bath's linearity is provided by probing the temperature dependence of the
spectral density,
which should be absent if the two are strictly fulfilled.
The results lead to similar conclusions, see Fig.~S7 in the Supplement.

In order to elucidate the impact of these non-linearities onto observables
we compare the GLE spectra with the ones obtained from explicit MD simulations in \Fig{fig:spectra}.
The GLE spectra (green curves therein) for aqueous systems (panels a,b) are significantly different from the reference ones 
(black curves therein) both in shape and position.
For the ionic liquid (panel c) these differences are not so pronounced, though still noticeable.
This comparison supports the  conjecture that the Gaussian statistics of the bath is more important,
although it cannot be considered as its proof in general.
Still, in all cases considered, the match of spectra is not achieved and thus the direct mapping $\Ad$ fails to exist for the systems studied.

The second route is based on the mapping $\Alp$, which is exact 
and therefore the corresponding spectra (red stars in \Fig{fig:spectra}) perfectly match the
explicit MD ones (black lines) for all systems considered.
Importantly, this mapping is established at the price of 
putting all the system  anharmonicity into the bath, thus giving an effectively harmonic \CL\ model with a frequency of $\tilde{\omega}$
(for an application to water, see e.g.~\cite{Fried1992}).
As a word of caution, we note that this model is tailored to yield linear response correctly,
whereas other observables  are not expected to be reproduced.

The third route is based on the mapping $\Anlp$, which employs the non-linear projection technique.
Note that although the derivation is exact, the memory kernel has been linearized to restore the desired equation structure.
The resulting spectra, orange curves in \Fig{fig:spectra}, improve slightly with respect to that of the direct 
mapping (green lines)
in terms of peak positions, but still have a wrong shape compared to the reference spectra (black lines).
Hence, the mapping $\Anlp$  is not sufficient to completely describe the spectra for the systems studied,
and the only successful approach is the effectively harmonic mapping $\Alp$.
As a side-result we note that intramolecular forces appear to be not very important, as it follows from the comparison of
panels a) and b) in \Fig{fig:spectra}.

To summarize, we have considered three ways to construct a mapping and have shown that only  $\Alp$, based on an effectively harmonic potential, gives correct spectra. However, such an effective harmonic system potential would not be suitable for describing nonlinear infrared spectroscopy since nonlinear response functions will vanish in this case. In principle, there could be a yet unknown mapping yielding anharmonic system potentials. To this end we would like to consider the problem from a more general perspective without being bound to a particular way of constructing the mapping. 
\begin{figure}[tb]
\begin{centering}
\includegraphics[width=0.99\columnwidth]{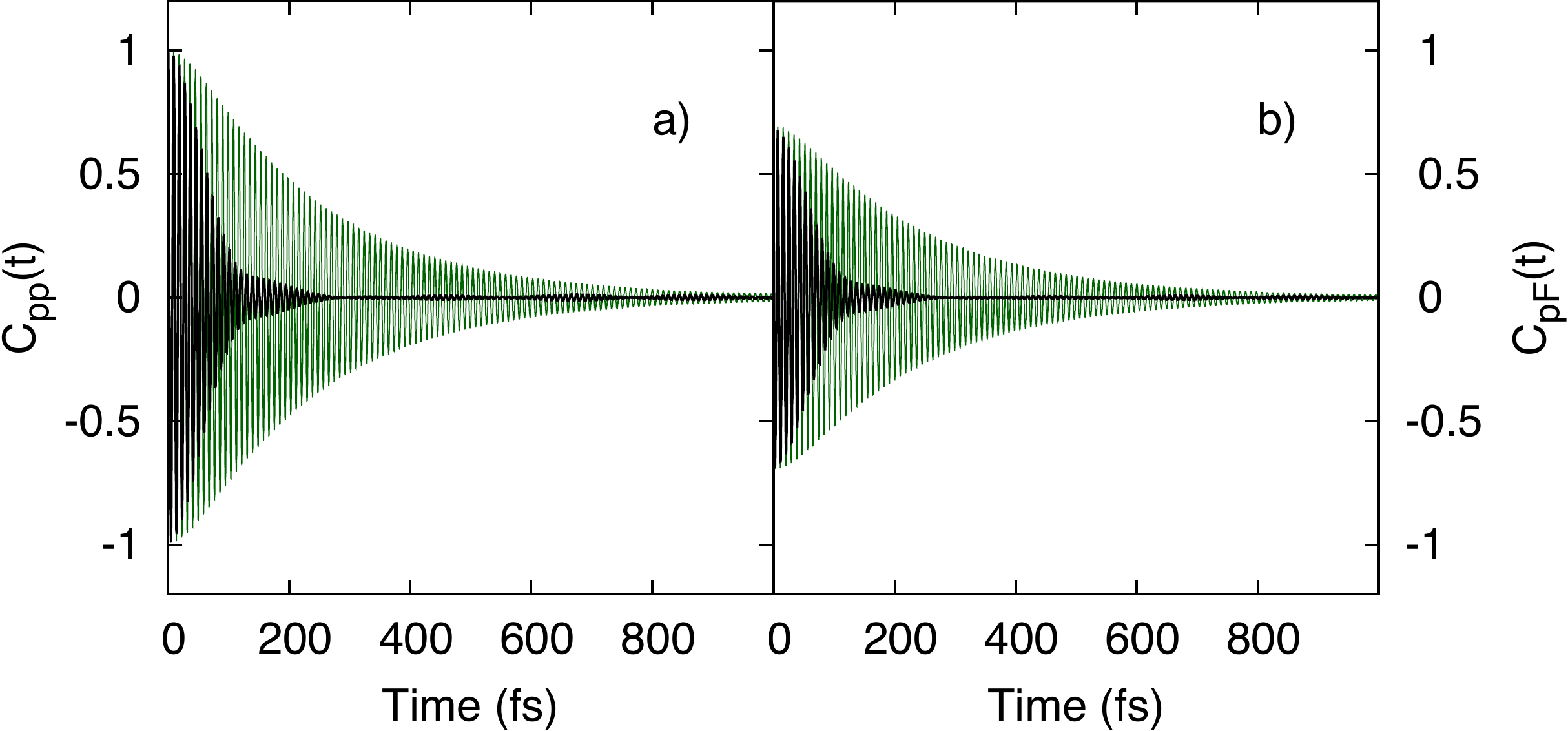}
\caption{\label{fig:uniqueness}
$C_{pp}$ in panel a) and $C_{pF}$ in panel b) are shown for the HOD in water case.
Black stands for explicit results, green -- for $\Ad$ according to \Eq{eq:GLE}.
}
\par\end{centering}
\end{figure}
We note that so far our assessment was based on vibrational spectra, that is on $C_{pp}^{}(t)$.
In fact, performing the dynamics according to \Eq{eq:GLE}
yields \textit{two} presumably independent correlation functions, and 
thus defines a mapping $\B: \xi(t)\mapsto \{C'_{pp}(t);C'_{pF}(t)\}$.
In order to have a self-consistent formalism  the pair of correlation functions $\{C'_{pp}(t)$;$C'_{pF}(t)\}$ should coincide with the pair $\{C_{pp}^{}(t)$;$C_{pF}^{}(t)\}$ obtained from the explicit MD simulations 
that was used to establish mapping $\A$.
Mathematically, we hence require $\B$ to be the inverse mapping of $\A$. 
Unfortunately, mapping $\A$ is not invertible, since it constitutes a mapping of two functions onto one,
unless the two functions are mutually dependent.
Thus, $\B$ cannot be the inverse mapping of $\A$ in the general case,
and the self-consistency requirement of the formalism is generally corrupted.
This problem is termed  the  ``invertibility problem''.

To illustrate this point on a particular example, the procedure described above led to two strikingly
\emph{different} pairs of $\{C_{pp}(t);C_{pF}(t)\}$ for mapping $\Ad$,
as is shown in \Fig{fig:uniqueness} for the case of HOD in water.
Importantly, in the case of the mapping $\Alp$ this problem does not surface,
since for a harmonic force $C_{pF}^{}(t)$ and $C_{pp}^{}(t)$ are mutually dependent as
$C_{pF}^{}(t)=-\tilde{\omega}^2\intop_{0}^{t} C_{pp}^{}(\tau) \diff \tau $.
%
Thus, the mapping $\Alp$ can be formulated with $C_{pp}^{}(t)$ alone and the invertibility problem does not occur.
This suggests that it is the anharmonicity of the system potential, $\VSt(x)$, in the GLE that is at the root of all deficiencies of the applicability of the \CL\ model to real anharmonic molecular systems.
Since this consideration is independent on  the particular form of the mapping $\A$,
the invertibility problem would surface in general, that is, for an arbitrary anharmonic system.

\paragraph{Conclusions and Outlook}
In this Letter the applicability of the \CL\ model to the dynamics of realistic anharmonic systems has been investigated.
Three possible mappings between a real system and the model, 
i.e.\ the direct mapping and the mappings based on (non-)linear projection techniques, have been considered.
Investigating three representative systems, we obtained
numerical evidence that only the mapping based on the linear projection technique ($\Alp$)
reproduces the linear absorption spectrum.
Importantly, this comes at the price of projecting the system's anharmonicity into the bath, resulting in an effectively harmonic potential which is not suitable for describing nonlinear spectroscopy.
The general argument against a mapping that leads to an \textit{anharmonic} system potential and gives a reasonable description have been provided and illustrated on the particular example of the direct mapping.
This argument, termed \textit{invertibility problem}, is not bound to the particular way of constructing the mapping and thus can be considered as a strong indication that such a mapping does not exist within the (bi-linear) \CL\ model.

One might ask whether extending the model by considering a non-linear coupling on the system side
would yield a solution.
Unfortunately, the structure of the GLE derived from such an extended model would differ from that obtained from
the rigorous projection techniques, see Supplement for a particular example of square-linear model.
%
This makes establishing the mapping for an extended model difficult if at all possible.

The authors gratefully acknowledge financial support by the Deutsche Forschungsgemeinschaft (Sfb 652). We thank Dr. Sergey Bokarev for fruitful discussions. 

\end{document}